\documentclass[twoside,leqno,twocolumn]{article}

\usepackage[letterpaper]{geometry}

\usepackage{siamproceedings}

\usepackage{amsmath,amsfonts}

\usepackage[noend]{algpseudocode}
\algrenewcommand\algorithmicthen{}

\makeatletter
\algnewcommand{\LineComment}[1]{\Statex \hskip\ALG@thistlm $\triangleright$ #1}
\makeatother

\usepackage{booktabs}
\usepackage{microtype}
\usepackage{graphicx}
\usepackage{textcomp}
\usepackage{xcolor}
\usepackage{xspace}
\usepackage{url}
\usepackage{balance}
\usepackage{tikz}
\usepackage{subcaption}
\usepackage{pgfplots}
\usepackage{multirow}
\usepackage{cleveref}
\usepackage{makecell}

\pgfplotsset{compat=1.9}
\def\BibTeX{{\rm B\kern-.05em{\sc i\kern-.025em b}\kern-.08em
    T\kern-.1667em\lower.7ex\hbox{E}\kern-.125emX}}

\usepackage{enumitem}
\setitemize{leftmargin=*,noitemsep,topsep=0pt,parsep=0pt,partopsep=0pt}
\setenumerate{leftmargin=*,noitemsep,topsep=0pt,parsep=0pt,partopsep=0pt}

\definecolor{magenta4}{rgb}{0.5625,0,0.5625}
\definecolor{green4}{rgb}{0,0.5625,0}
\definecolor{orange4}{rgb}{0.98,0.31,0.09}

\definecolor{o-bskip}{HTML}{B3BCE9}
\definecolor{bskip}{HTML}{BCEAF2}
\definecolor{folly}{HTML}{C7E0BC}
\definecolor{btree}{HTML}{FAE2BE}
\definecolor{olc-btree}{HTML}{F9BAB3}

\definecolor{o-bskip-dark}{HTML}{6168CF}
\definecolor{bskip-dark}{HTML}{75ACD6}
\definecolor{folly-dark}{HTML}{79AD69}
\definecolor{btree-dark}{HTML}{CF9771}
\definecolor{olc-btree-dark}{HTML}{C96060}

\newcommand{\helen}[1]{{ \textcolor{magenta4}{Helen: {#1}}}}

\newcommand{\figref}[1]         {Figure~\ref{fig:#1}}

\newcommand{\tabref}[1]        {Table~\ref{tab:#1}}
\newcommand{\secref}[1]         {Section~\ref{sec:#1}}

\newcommand{\numthreads}{\ensuremath{128}\xspace}

\newcommand{\skiplist}{skiplist\xspace}
\newcommand{\skiplists}{skiplists\xspace}
\newcommand{\bskiplist}{B-\skiplist}
\newcommand{\bskipnospace}{\bskiplist}
\newcommand{\bskip}{\bskiplist\xspace}
\newcommand{\bskiplists}{B-\skiplists}

\newcommand{\btree}{B-tree\xspace}
\newcommand{\bplustree}{B$^+$-tree\xspace}
\newcommand{\bplustreenospace}{B$^+$-tree}

\newcommand{\obskip}{optimistic \bskip}
\newcommand{\obskipnospace}{optimistic \bskipnospace}

\newcommand{\rwbased}{RW-lock-based\xspace}
\newcommand{\olcbased}{OLC-based\xspace}
\newcommand{\olcbskip}{OLC-\bskipnospace}
\newcommand{\olcbtree}{OLC-\bplustreenospace}
\newcommand{\rwbskip}{RW-\bskipnospace}
\newcommand{\rwbtree}{RW-\bplustreenospace}

\newcommand{\obskipplot}{OLC-BSL\xspace}
\newcommand{\rwbskipplot}{RW-BSL\xspace}
\newcommand{\follyplot}{FLY-SL\xspace}
\newcommand{\rwbtreeplot}{RW-B$^+$T\xspace}
\newcommand{\olcbtreeplot}{OLC-B$^+$T\xspace}

\newcommand{\btrees}{B-trees\xspace}

\newcommand{\maxlen}{\texttt{max\_len}\xspace}
\newcommand{\defn}[1]{\textit{#1}}

\newcommand{\folly}{Folly\xspace}

\renewcommand{\defn}[1]{{\textit{\textbf{#1}}}}

\newcommand{\obsloverslthroughputrange}{$1.2$--$10\times$\xspace}

\newcommand{\obskipoverbtreethroughputrange}{$0.8$--$4.3\times$\xspace}

\newcommand{\para}[1]{\smallskip\noindent\textbf{#1.}}

\usetikzlibrary{calc,matrix, shapes.multipart,patterns}
\pgfplotsset{
    discard if/.style 2 args={
        x filter/.append code={
            \edef\tempa{\thisrow{#1}}
            \edef\tempb{#2}
            \ifx\tempa\tempb
                
            \fi
        }
    },
    discard if not/.style 2 args={
        x filter/.append code={
            \edef\tempa{\thisrow{#1}}
            \edef\tempb{#2}
            \ifx\tempa\tempb
            \else
                
            \fi
        }
    },
}

\pgfplotsset{compat=1.8,
    /pgfplots/ybar legend/.style={
    /pgfplots/legend image code/.code={%
       \draw[##1,/tikz/.cd,yshift=-0.25em]
        (0cm,0cm) rectangle (3pt,0.8em);},
   },
}
\definecolor{safe-black}{RGB}{0,0,0}
\definecolor{safe-olive}{RGB}{0,73,73}
\definecolor{safe-teal}{RGB}{0,146,146}
\definecolor{safe-pink}{RGB}{255,109,182}
\definecolor{safe-peach}{RGB}{255,160,110}
\definecolor{safe-plum}{RGB}{73,0,146}
\definecolor{safe-cerulean}{RGB}{0,109,219}
\definecolor{safe-lavender}{RGB}{182,109,255}
\definecolor{safe-sky}{RGB}{109,182,255}
\definecolor{safe-baby}{RGB}{182,219,255}
\definecolor{safe-brick}{RGB}{146,0,0}
\definecolor{safe-brown}{RGB}{146,73,0}
\definecolor{safe-orange}{RGB}{219,209,0}
\definecolor{safe-green}{RGB}{36,255,36}
\definecolor{safe-yellow}{RGB}{255,255,109}

\pgfplotscreateplotcyclelist{mylist}{%
    {safe-pink, mark=square*},
    {safe-cerulean, mark=triangle*},
    {safe-peach, mark=x,mark options={scale=1.6}},
    {blue, mark=star,mark options={scale=1.6}},
    }

    \pgfplotscreateplotcyclelist{mylistshapes}{%
    {safe-pink, mark=square*},
    {safe-cerulean, mark=triangle*},
    {safe-peach,mark=*},
    {blue, mark=diamond*},
    }

\newcommand{\VALIDATE}[2]{\textsc{Validate}(#1,\, #2)}
\newcommand{\IsWriteLocked}[1]{\textsc{IsWriteLocked}(#1)}
\newcommand{\LOADVERSION}[1]{\textsc{LoadVersion}(#1)}

\newcommand{\tpAll}{0.8--10$\times$\xspace}

\newcommand{\tpWrite}{1.2--4.4$\times$\xspace}

\newcommand{\tpBskipL}{2.7$\times$\xspace}
\newcommand{\tpFollyL}{4$\times$\xspace}
\newcommand{\tpRWBTreeL}{4.3$\times$\xspace}
\newcommand{\tpOLCBTreeL}{3$\times$\xspace}

\newcommand{\tpBskipABC}{1.2--1.8$\times$\xspace}
\newcommand{\tpFollyABC}{3.7--4.4$\times$\xspace}
\newcommand{\tpRWBTreeABC}{1.1--2.4$\times$\xspace}
\newcommand{\tpOLCBTreeABC}{0.8--1.2$\times$\xspace}

\newcommand{\tpBskipE}{1.3--1.4$\times$\xspace}
\newcommand{\tpFollyE}{9--10$\times$\xspace}
\newcommand{\tpRWBTreeE}{1.2$\times$\xspace}

\newcommand{\latHigher}{1.6$\times$\xspace}
\newcommand{\latMax}{13$\times$\xspace}
\newcommand{\latBskipL}{1.2--13$\times$\xspace}
\newcommand{\latFollyL}{2.8--3.7$\times$\xspace}
\newcommand{\latRWBTreeL}{1.3--13$\times$\xspace}

\newcommand{\latHigherOLCBTreeL}{1.3--1.4$\times$\xspace}

\newcommand{\latBskipABC}{0.9--1.7$\times$\xspace}
\newcommand{\latFollyABC}{2.6--3.8$\times$\xspace}
\newcommand{\latRWBTreeABC}{0.8--6$\times$\xspace}

\newcommand{\latBskipE}{0.9--1.2$\times$\xspace}
\newcommand{\latFollyE}{8--12$\times$\xspace}
\newcommand{\latRWBTreeE}{0.7--2.9$\times$\xspace}

\newcommand{\olcbskipscalingA}{74$\times$\xspace}
\newcommand{\olcbskipscalingC}{81$\times$\xspace}
\newcommand{\olcbskipscalingL}{57$\times$\xspace}

\newcommand{\rwbskipscalingA}{45$\times$\xspace}
\newcommand{\rwbskipscalingC}{70$\times$\xspace}
\newcommand{\rwbskipscalingL}{22$\times$\xspace}

\newcommand{\follyscalingA}{68$\times$\xspace}
\newcommand{\follyscalingC}{88$\times$\xspace}
\newcommand{\follyscalingL}{60$\times$\xspace}

\newcommand{\rwbtreescalingA}{34$\times$\xspace}
\newcommand{\rwbtreescalingC}{74$\times$\xspace}
\newcommand{\rwbtreescalingL}{12$\times$\xspace}

\newcommand{\olcbtreescalingA}{43$\times$\xspace}
\newcommand{\olcbtreescalingC}{64$\times$\xspace}
\newcommand{\olcbtreescalingL}{15$\times$\xspace}

\usepackage{amsopn}
\DeclareMathOperator{\diag}{diag}

\begin{document}

\newcommand\relatedversion{}
\renewcommand\relatedversion{\thanks{The full version of the paper can be accessed at \protect\url{https://arxiv.org/abs/0000.00000}}} 

\title{A Simple Optimistic B-Skiplist Algorithm}
\author{Jianting Guo\thanks{\raggedright\hangindent=1.5em\hangafter=1
  National Chung Hsing University
  (\email{ting1116@smail.nchu.edu.tw}).}
  \and Yicong Luo\footnotemark[2]
  \and Helen Xu\thanks{\raggedright\hangindent=1.5em\hangafter=1
  Georgia Institute of Technology
  (\email{yluo460@gatech.edu}, \email{hxu615@gatech.edu}).}}
\date{}

\maketitle


\fancyfoot[R]{\scriptsize{Copyright \textcopyright\ 2027 by SIAM\\
Unauthorized reproduction of this article is prohibited}}





\begin{abstract}
  Skiplists underlie the in-memory index in many key-value stores, such as
  RocksDB and LevelDB, due to their implementation simplicity and
  straightforward concurrency control. Traditional skiplists exhibit poor cache
  locality since they store only one element per node. Prior work narrowed this
  gap with a concurrent B-skiplist that stores multiple elements per node and
  uses reader-writer (RW) locks for one-pass concurrency. However, the
  state-of-the-art concurrent B-skiplist relies on partitioned RW locks that
  span multiple cache lines, resulting in excess lock-induced cache misses.

  We present the first optimistic concurrent B-skiplist, built on optimistic
  lock coupling (OLC). OLC replaces the multi-cache-line RW lock with a compact,
  single-cache-line lock: readers validate a version number without writing to
  shared memory, avoiding cache invalidations and retrying only on a detected
  conflict. Although OLC was thought poorly suited to skiplists, where a single
  per-element lock may span structurally unrelated levels, the B-skiplist's
  blocked layout places each level (and associated lock) in its own node. We
  exploit the B-skiplist's upfront knowledge of promotion height with a hybrid
  top-down scheme that traverses upper levels optimistically and locks only from
  the promotion level down, so most inserts never dirty a cache line.

  On \numthreads threads, the proposed \obskip achieves
  \obsloverslthroughputrange higher throughput over state-of-the-art
  concurrent skiplist implementations. Furthermore, the \obskip achieves
  \obskipoverbtreethroughputrange the throughput compared to high-performance
  optimistic \bplustree implementations based on RW locks and OLC. Finally, the \obskip achieves up to $13\times$ lower latency than
  concurrent skiplists and tree-based indices, and is within $1.6\times$ of the
  latency of the OLC-based \bplustree.

  The artifact is made available at~\cite{artifact}.
\end{abstract}


\section{Introduction.}\label{sec:intro}

\begin{table}[t]
\centering
\small
\setlength{\tabcolsep}{5pt}
\begin{tabular}{@{}lcccc@{}}
\toprule
\multirow{2}{*}{\bskiplist} & \multicolumn{2}{c}{Cache} & Memory & Throughput \\
\cmidrule(lr){2-3}
 & Misses & References & (GB) & (ops/$\mu$s) \\
\midrule
RW-lock    & 61E9 & 220E9 & 6.7 & 59 \\
Optimistic & 57E9 & 120E9 & 6.3 & 102 \\
\bottomrule
\end{tabular}
\caption{Cache misses, cache references, memory usage, and throughput of the
  optimistic \bskiplist (this work) and the RW-lock \bskiplist~\cite{bskip}.
  Throughput is for YCSB~\cite{ycsb} workload A (50\% inserts/50\% finds);
  all other data is for load + workload A (200M ops total).}
\label{tab:intro_table}
\end{table}

The \skiplist~\cite{pugh1990skip} is a popular choice for the in-memory index
(i.e., the memtable) in many key-value stores, such as HBase~\cite{hbase},
RocksDB~\cite{rocksdb}, and LevelDB~\cite{LevelDB}, for its simple structure and
concurrency control (CC) schemes~\cite{herlihy2006provably, pugh1990concurrent,
  herlihy2007simple, fomitchev2004lock, fraser2004practical}. The \skiplist
stores elements in a collection of ordered linked lists. The bottom level
contains all the key-value pairs, and each subsequent level contains a subset of
the elements of the level immediately below it. The \skiplist determines which
levels a key appears in \emph{probabilistically}, avoiding complex rebalancing
operations.


The simple structure of the \skiplist comes at the cost of poor locality, since
each node stores only one key. Blocked data structures like the
\btree~\cite{BayerMc72} alleviate this by collocating keys in a single
node, reducing cache misses during traversal. In the I/O model (or
external-memory model)~\cite{AggarwalVi88}, blocking nodes asymptotically
improves the height (and traversal time) of a \btree compared to a
\skiplist or binary tree. Given a cache-line size $Z$, the operation time (find
and insert) of a \btree is $O(\log_Z n)$, asymptotically smaller than the
$O(\log_2 n)$ time of a \skiplist or binary tree.

\para{Bridging locality and concurrency} Luo \emph{et al.}~\cite{bskip}
introduced a concurrent \defn{B-skiplist} that collocates elements within a
node. Rather than promoting elements with probability $1/2$, a \bskiplist with node size $B$
promotes elements with probability $\Theta(1/B)$ and, for all levels
$\ell = 0, 1, \ldots, \log_B n$, collocates elements on level $\ell$ between
elements promoted to level $\ell + 1$. Golovin~\cite{golovin2010b} showed
theoretically that when $B = \Theta(Z)$, the \bskiplist matches the bounds of the classical \btree in
expectation, and these results are borne out in practice~\cite{bskip}: the
\bskiplist achieves over $2\times$ higher throughput and $3$--$5\times$ fewer cache
misses than a state-of-the-art concurrent \skiplist from Facebook's Folly
library~\cite{folly}.

\para{Locality issues in today's concurrent \bskiplists} To support concurrent
operations, the \bskiplist employs \defn{partitioned reader-writer (RW)
  locks}~\cite{courtois1971concurrent} with a one-pass traversal algorithm that
avoids revisiting nodes~\cite{bskip}. RW locks can be acquired in \defn{shared}
mode (by incrementing a reader counter) or in \defn{exclusive} mode (by setting
a write bit). Naive RW lock implementations do not scale well because any
write to a shared-memory location invalidates copies of the cache line across
all cores, even when the lock is acquired in shared mode. The RW locks in the
\bskiplist therefore use partitioning~\cite{paralleltools}
with multiple reader counts (one per cache line) to avoid false sharing.

\begin{figure}[t]
  \centering
  \begin{tikzpicture}
    \begin{axis}[
    width=7.8cm, height=5cm,
        ybar,
        legend style={at={(0.4, 1.4)}, anchor=north},
        ylabel={Throughput (ops/$\mu$s)},
        ylabel style={align=center, text width=3cm},
        xlabel={Workload},
        xtick=data,
        bar width=4pt,
        legend columns=5,
        legend style={font=\small},
        ytick={40,80,120,160},
        enlarge x limits=0.14,
        symbolic x coords={Load, A, B, C, E},
        x tick label style={text width = 1cm, align = center},
        ymin = 0,
        ymax = 180,
      ]

    \addplot [fill=o-bskip, draw=o-bskip-dark!70, thick, forget plot]
    coordinates {
        (Load,96.94)(A, 102.91)(B, 124.28)(C, 130.59)(E, 86.7)
    };
    \addplot [
        fill=o-bskip,
        draw=o-bskip-dark,
        pattern=none,
        pattern color=o-bskip-dark!70,
        legend image code/.code={
            \draw[fill=o-bskip, draw=o-bskip-dark!70, thick] (-0.07cm,-0.15cm) rectangle (0.07cm,0.15cm);
            \draw[pattern=none, pattern color=o-bskip-dark!70, draw=none] (-0.07cm,-0.15cm) rectangle (0.07cm,0.15cm);
        }
    ]
    coordinates {
        (Load,96.94)(A, 102.91)(B, 124.28)(C, 130.59)(E, 86.7)
    };
    \addlegendentry{\obskipplot}

    \addplot [fill=bskip, draw=bskip-dark, thick, forget plot]
    coordinates {
        (Load,36.47)(A, 59.15)(B, 103.28)(C, 107.43)(E, 65.19)
    };
    \addplot [
        fill=bskip,
        draw=bskip-dark,
        pattern=crosshatch dots,
        pattern color=bskip-dark!80,
        legend image code/.code={
            \draw[fill=bskip, draw=bskip-dark, thick] (-0.07cm,-0.15cm) rectangle (0.07cm,0.15cm);
            \draw[pattern=crosshatch dots, pattern color=bskip-dark!80, draw=none] (-0.07cm,-0.15cm) rectangle (0.07cm,0.15cm);
        }
    ]
    coordinates {
        (Load,36.47)(A, 59.15)(B, 103.28)(C, 107.43)(E, 65.19)
    };
    \addlegendentry{\rwbskipplot}

    \addplot [fill=folly, draw=folly-dark, thick, forget plot]
    coordinates {
        (Load,24.40)(A, 24.81)(B, 33.59)(C, 35.35)(E, 9.51)
    };
    \addplot [
        fill=folly,
        draw=folly-dark,
        pattern=crosshatch,
        pattern color=folly-dark!80,
        legend image code/.code={
            \draw[fill=folly, draw=folly-dark, thick] (-0.07cm,-0.15cm) rectangle (0.07cm,0.15cm);
            \draw[pattern=crosshatch, pattern color=folly-dark!80, draw=none] (-0.07cm,-0.15cm) rectangle (0.07cm,0.15cm);
        }
    ]
    coordinates {
        (Load,24.40)(A, 24.81)(B, 33.59)(C, 35.35)(E, 9.51)
    };
    \addlegendentry{\follyplot}

    \addplot [fill=btree, draw=btree-dark, thick, forget plot]
    coordinates {
        (Load,22.53)(A, 47.55)(B, 101.33)(C, 119.84)(E, 72.98)
    };
    \addplot [
        fill=btree,
        draw=btree-dark,
        pattern=north east lines,
        pattern color=btree-dark!80,
        legend image code/.code={
            \draw[fill=btree, draw=btree-dark, thick] (-0.07cm,-0.15cm) rectangle (0.07cm,0.15cm);
            \draw[pattern=north east lines, pattern color=btree-dark!80, draw=none] (-0.07cm,-0.15cm) rectangle (0.07cm,0.15cm);
        }
    ]
    coordinates {
        (Load,22.53)(A, 47.55)(B, 101.33)(C, 119.84)(E, 72.98)
    };
    \addlegendentry{\rwbtreeplot}


    \addplot [fill=olc-btree, draw=olc-btree-dark, thick, forget plot]
    coordinates {
        (Load, 32.5)(A, 86.1)(B, 144.8)(C, 158)(E, 0)
    };
    \addplot [
        fill=olc-btree,
        draw=olc-btree-dark,
        pattern=north west lines,
        pattern color=btree-dark!80,
        nodes near coords,
        nodes near coords style={
            rotate=90,
            anchor=west,
            font=\footnotesize,
        },
        point meta=explicit symbolic,
        legend image code/.code={
            \draw[fill=olc-btree, draw=olc-btree-dark, thick] (-0.07cm,-0.15cm) rectangle (0.07cm,0.15cm);
            \draw[pattern=north west lines, pattern color=olc-btree-dark!80, draw=none] (-0.07cm,-0.15cm) rectangle (0.07cm,0.15cm);
        }
    ]
    coordinates {
        (Load, 32.5)(A, 86.1)(B, 144.8)(C, 158)(E, 0)[\texttt{N/A}]
    };
    \addlegendentry{\olcbtreeplot}


    \end{axis}
  \end{tikzpicture}
  \vspace{-.5cm}
  \caption{Throughput on uniform YCSB workloads on 128 threads. The load phase
    and workloads A--C involve varying ratios of point finds and inserts, while
    workload E involves range queries and inserts.}
    \label{fig:tp_uniform}
  \end{figure}
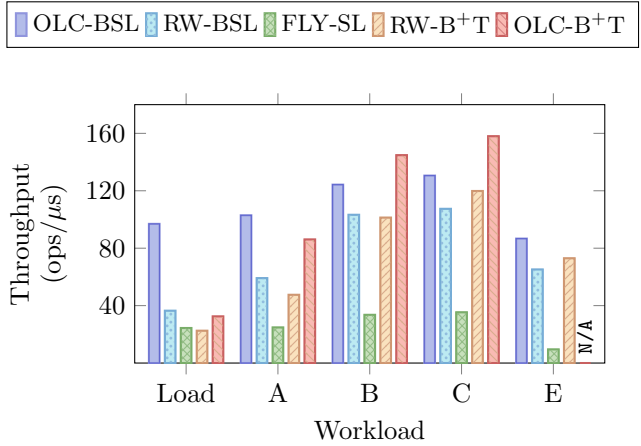

However, partitioning introduces new locality challenges: writers must scan all
partitions to acquire the lock, incurring additional cache misses. For example,
with 128 threads and 64-byte partitions (1 cache line each), the partitioned
counter is 8192 bytes---far larger than the default \bskiplist node size of 2048
bytes~\cite{bskip}. The concurrency-control metadata thus adds cache-line accesses to every node
visit, eroding the locality benefits of the B-skiplist's blocked layout.

\para{Our approach: optimistic lock coupling}
\emph{We introduce a concurrent \bskiplist based on
  \defn{optimistic lock coupling} (OLC)~\cite{leis2019optimistic} that improves
  the cache locality and performance of the state-of-the-art concurrent
  \bskiplist by replacing its partitioned RW lock with a compact, cache-friendly
  OLC lock. The proposed \obskip maintains the simplicity of traditional
  \skiplist CC schemes while improving performance and scalability.}

OLC~\cite{leis2019optimistic} improves cache performance by combining lock-based
and lock-free synchronization: readers validate a version number without writing
to shared memory, avoiding cache invalidations, while the entire lock fits on a
single cache line. In B-trees, Leis \emph{et al.}  demonstrated that OLC
achieves near-linear scalability, outperforming traditional lock coupling by up
to $4\times$~\cite{leis2019optimistic}.

To our knowledge, this is the first application of OLC to a B-skiplist.  OLC has
been applied to linked lists~\cite{guerraoui2016optimistic},
\skiplists~\cite{guerraoui2016optimistic}, binary
trees~\cite{bronson2010practical}, and B-trees~\cite{cha2001cache,
  leis2019optimistic}.

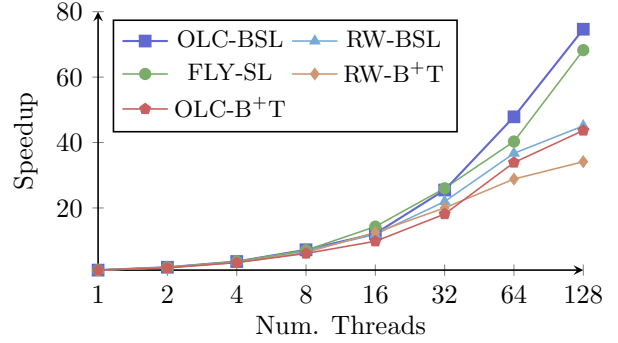
\begin{figure}[t]
  \centering
  \begin{tikzpicture}
    \begin{axis}[
        width=8cm, height=5cm,
        axis lines = left,
        xlabel = Num. Threads,
        ylabel style={align=center},
        ylabel = {Speedup},
        cycle list name=exotic,
        legend pos=north west,
        legend style={font=\small},
        xmode=log,
        log basis x={2},
        legend columns=2,
        ymax=80,
        ytick={20,40,60,80},
        log ticks with fixed point,
        xlabel shift={-2pt},
        xtick={1,2,4,8,16,32,64,128},
        xticklabels = {1,2,4,8,16,32,64,128},
        semithick,
      ]
      \addplot[mark=square*, color=o-bskip-dark, thick]
      coordinates{
        (1,   1)
        (2,   1.873513572)
        (4,   3.626538287)
        (8,   7.21803913)
        (16,  12.15564743)
        (32,  25.48129758)
        (64,  47.86407659)
        (128, 74.64)
      };
      \addlegendentry{\obskipplot}

    \addplot[mark=triangle*, color=bskip-dark]
      coordinates{
        (1,   1)
        (2,   1.83)
        (4,   3.53)
        (8,   6.89)
        (16,  11.92)
        (32,  21.96)
        (64,  36.71)
        (128, 45.09)
      };
      \addlegendentry{\rwbskipplot}
      \addplot[mark=*, color=folly-dark]
      coordinates{
        (1,   1)
        (2,   1.84)
        (4,   3.60)
        (8,   7.08)
        (16,  14.29)
        (32,  26.01)
        (64,  40.31)
        (128, 68.24)
      };
      \addlegendentry{\follyplot}

      \addplot[mark=diamond*, color=btree-dark]
      coordinates{
        (1,   1)
        (2,   1.77)
        (4,   3.33)
        (8,   6.37)
        (16,  12.50)
        (32,  20.02)
        (64,  28.83)
        (128, 34.15)
      };
      \addlegendentry{\rwbtreeplot}

      \addplot[mark=pentagon*, color=olc-btree-dark]
      coordinates{
        (1,   1)
        (2,   1.69)
        (4,   3.28)
        (8,   6.09)
        (16,  9.81)
        (32,  18.13)
        (64,  33.88)
        (128, 43.68)
      };
      \addlegendentry{\olcbtreeplot}

    \end{axis}
  \end{tikzpicture}
    \vspace{-.3cm}
    \caption{Strong scaling on uniform YCSB workload A (50\%/50\% finds/inserts).}
  \label{fig:scaling_a}
\end{figure}

\para{Challenges to OLC in the \bskiplist} Prior work on OLC-based
\skiplists~\cite{guerraoui2016optimistic} observed that OLC is poorly suited to
skiplists, because a skiplist typically stores an element's forward pointers for
all its levels in a single tower under one per-element lock, coupling
unrelated levels. The tower optimization is meant to improve locality in
\skiplists by avoiding cache misses during downwards traversals, but comes at
the cost of coarser lock granularity producing false conflicts.

The \bskiplist optimizes for locality without giving up on concurrency through
its blocked layout: since each block collocates different elements at each
level, promoted elements must reside in a different node at each level.
Therefore, the \bskiplist's blocking also enables fine-grained, per-level
concurrency.

Finally, applying OLC to the \bskiplist raises additional challenges not present
in B-trees because 1) \bskiplist traversals require both horizontal and vertical
steps, requiring version reads and validations in two dimensions rather than
one, and 2) promoted inserts must update forward pointers at multiple levels,
requiring careful ordering of pointer updates and version increments to avoid
blocking readers unnecessarily.

\para{Designing an efficient OLC-based \bskiplist}
 Despite these challenges, the B-skiplist's structure and randomized promotion make it well-suited for OLC. Building on the top-down insertion algorithm of the RW-lock-based B-skiplist (details in~\secref{prelim}), our CC scheme replaces the multi-cache-line RW lock with a single-cache-line OLC lock, incurring $1.8\times$ fewer cache references on write-heavy workloads (\tabref{intro_table}). 
 
 We further exploit the B-skiplist's upfront knowledge of promotion height with a \defn{hybrid} top-down mechanism: writers traverse levels above the promotion height optimistically, never acquiring a write lock, and take write locks only at the promotion level and below. Since promotion probability decreases geometrically, most inserts reach low heights, so the write lock covers only a few levels and writers access the vast majority of nodes without dirtying a cache line. When a retry does occur, it is cheap, as the traversal path is likely still in cache.

\para{Contributions} This paper makes the following contributions:
\begin{itemize}[leftmargin=*,noitemsep]
\item The design of an \olcbased concurrent \bskiplist that optimizes for
  locality in the concurrency control mechanism.
\item A \emph{hybrid} top-down \olcbased CC mechanism for the \bskiplist that
  completes insertions in a single top-to-bottom pass.
\item A C implementation of the \olcbased concurrent \bskiplist.
\item An empirical evaluation against several in-memory skiplist-based and
  B-tree-based indices
  on YCSB~\cite{ycsb} and TPC-C~\cite{leutenegger1993modeling} workloads.
\end{itemize}

\para{Results summary} We evaluate the \olcbased \bskiplist (\obskipplot)
against the \rwbased \bskiplist (\rwbskipplot)~\cite{bskip}, Facebook's \folly
ConcurrentSkipList (\follyplot)~\cite{folly}, an optimized \rwbased \bplustree
(\rwbtreeplot)~\cite{XuLiWh23}, and an \olcbased \bplustree
(\olcbtreeplot)~\cite{leis2019optimistic, WangPaLi18} on YCSB~\cite{ycsb}
workloads (details in~\secref{eval}).
\figref{tp_uniform} shows throughput across workloads; \figref{scaling_a} shows
strong scaling on workload A.

On 128 threads, the \obskip achieves \tpAll the throughput and from \latHigher higher to \latMax lower
latency when compared to all other tested systems. As shown in~\figref{tp_uniform}, on the
write-heavy workloads (load and workload A), the \olcbskip achieves \tpWrite
speedup compared to the other indices. Furthermore, on the read-heavy point
workloads B and C, the \olcbskip achieves higher ($1.2$--$4\times$) speedup
compared to \skiplist-based indices and is competitive ($0.8$--$0.9\times$) with
the \olcbtree, which is the fastest tested \btree-based index.  In terms of
scaling, the \obskip achieves \olcbskipscalingA, \olcbskipscalingC, and
\olcbskipscalingL throughput speedup over 1 thread on
YCSB~\cite{ycsb-core-workloads} workloads A and C, and the load phase,
respectively.


\section{Background.}\label{sec:prelim}

This section reviews the structure and operations of skiplists and \bskiplists,
then covers the concurrency primitives this paper builds on: lock coupling,
reader-writer (RW) locks, and optimistic lock coupling (OLC).

\para{Operations} 
A key-value dictionary stores key-value pairs \texttt{(k, v)} with the
following operations:
\begin{itemize}
\item \texttt{find(k)}: return the associated value \texttt{v}.
\item \texttt{insert(k, v)}: add \texttt{(k, v)} to the data
  structure.
\item \texttt{delete(k)}: remove \texttt{k} and its associated value.
\item \texttt{range(k, f, length)}: apply the function \texttt{f} to the
  \texttt{length} key-value pairs with the smallest keys that are at least
  \texttt{k}.
\end{itemize}
Finds, inserts, and range queries comprise the YCSB~\cite{ycsb}
workloads used in our evaluation (\secref{eval}).

\para{Skiplist structure and operations}
The \skiplist~\cite{pugh1990skip} is a
\emph{hierarchy} of ordered linked lists: level 0 contains all keys, and each
level $\ell > 0$ contains a subset of the elements at level $\ell-1$, bounded by
a left sentinel ($-\infty$) and a right sentinel ($\infty$).
As in trees, nodes at level $\ell = 0$ are \defn{leaf} nodes and those at levels
$\ell > 0$ are \defn{internal} nodes. All nodes have a \texttt{next} pointer to
their successor in the same level; internal nodes also have a \texttt{down}
pointer to the node with the same key in the level below, and the \defn{root} is
the leftmost node at the highest level.

The \skiplist's defining feature is its \emph{randomized} self-balancing:
each element's height equals the number of successive ``heads'' when flipping a
coin with probability $p$ (usually $p = 1/2$; any constant $p = 1/c$ suffices for
the asymptotic bounds). Thus the maximum height of an element in a \skiplist with
$n$ elements is $O(\log n)$ in expectation and with high probability (w.h.p.), and
an element's promotion level is independent of the \skiplist structure.
At each level, the \texttt{pred} (resp.\ \texttt{succ}) of a key \texttt{k} is
the largest key less than \texttt{k} (resp.\ smallest greater than \texttt{k}).

All \skiplist operations traverse left-to-right and top-to-bottom from the root,
supporting point operations in $O(\log n)$ time in expectation and w.h.p.; we
detail the blocked variant in \secref{bskiplist}.

\subsection{\bskiplist structure and operations.}\label{sec:bskiplist}

\begin{figure}[t]
  \centering \includegraphics[width=\linewidth]{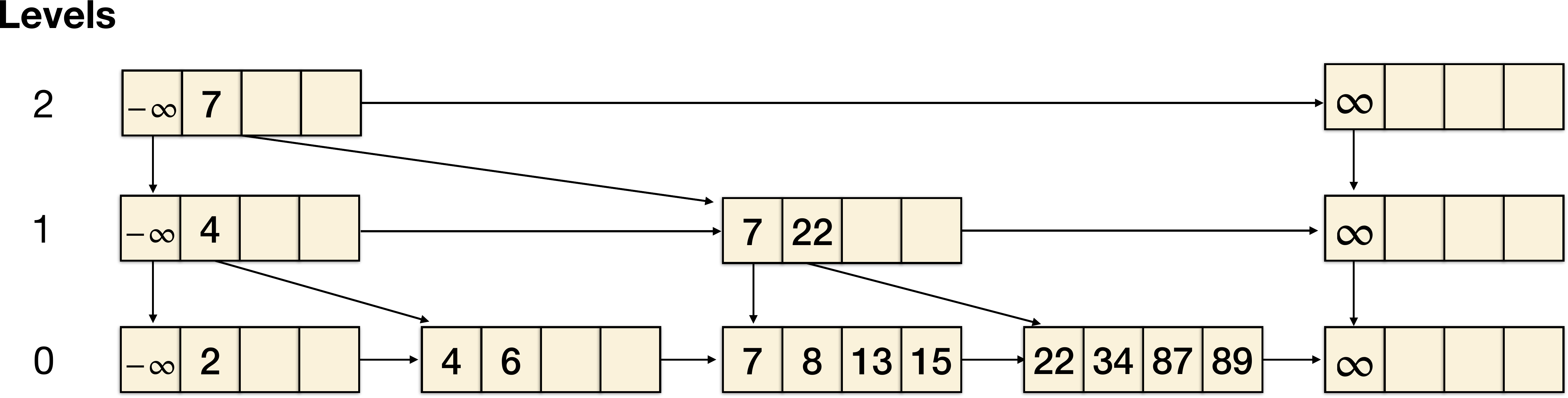}
  \vspace{-.75cm}
  \caption{Example of a \bskiplist with node size $B = 4$.}
  \label{fig:bskip-example}
\end{figure}

\para{Structure} The \bskiplist~\cite{golovin2010b} is a blocked \skiplist with
multiple elements per node. 
Given a cache-line size $Z$ and node size $B = \Theta(Z)$, the promotion
probability is $p = \Theta(1/B)$.

The \defn{header} of a node is its first (and smallest) key; by construction,
every header at level $\ell$ has been promoted to level $\ell + 1$.

Each
node in a \bskiplist has a \texttt{next} pointer, and all keys in node \texttt{x} are smaller
than the header of \texttt{x->next}. Each internal node (level $\ell > 0$) also
contains an array of $B$ \texttt{down} pointers, one per key, each pointing to
the corresponding node at the level below.

\para{Bounds} In the external-memory model (\secref{intro}), a \bskiplist with
$n$ elements and node size $B = \Theta(Z)$ supports finds and inserts in
$\Theta(\log_Z(n))$ cache-line transfers in expectation, matching B-tree bounds.
Range queries over $r$ elements take $\Theta(\log_Z(n) + r/Z)$ transfers in
expectation.

\para{Operations} \bskiplist traversals follow the same left-to-right,
top-to-bottom order as the \skiplist, but examine multiple elements per node.

The \texttt{find(k)} operation initializes \texttt{curr} to the upper-left
sentinel and scans right until the header of \texttt{curr->next} exceeds
\texttt{k}, then follows the \texttt{down} pointer of the \texttt{pred} element
within \texttt{curr}.  This repeats until the leaf level, where it checks for
\texttt{k}'s presence.  Range queries perform a \texttt{find} to the start key,
then iterate left-to-right at the leaf level.

The \texttt{insert(k, v)} operation performs a \texttt{find}-like traversal to
locate the \texttt{pred} at each level. Let $h$ be the promotion height of
\texttt{k}. If $h = 0$, \texttt{k} is inserted into the leaf node containing its
\texttt{pred}.  If $h > 0$, a \texttt{split} is performed at levels
$\ell = 0, \ldots, h - 1$, linking in new nodes with \texttt{k} as the header;
at each level, keys greater than \texttt{k} are copied to the new node.

Deletes are symmetric to inserts, and disconnect and deallocate a node when deleting the last key in the node.

\para{Insertion order in a \bskiplist} Golovin's theoretical paper~\cite{golovin2010b}
does not prescribe the insertion order (top-down or bottom-up). Most skiplist
algorithms use \defn{bottom-up} insertion, linking new nodes from the leaf level
up to the promotion height~\cite{pugh1990skip, pugh1990concurrent,
  herlihy2006provably, herlihy2007simple}. The RW-lock-based
\bskiplist~\cite{bskip} instead uses \defn{top-down} insertion, starting from
the promotion height down to the leaf level. The key advantage of this design is that insertion can finish in one top-down traversal compared to a bottom-up algorithm, which requires a top-down traversal followed by a bottom-up traversal.

\subsection{Concurrency control.}

\begin{algorithm}[t]
\caption{Lock coupling step}
\label{alg:hoh}
\begin{algorithmic}[1]
\State \textbf{Input:} $x$: current node, read lock of $x$ held from previous iteration
\State $y \gets x.next$
\State $\text{read\_lock}(y)$
\State $\text{read\_unlock}(x)$
\State $x \gets y$
\end{algorithmic}
\end{algorithm}

\para{Lock coupling} \defn{Lock coupling} (also known as \emph{hand-over-hand}
locking, HOH)~\cite{bayer1977concurrency, graefe2011modern, herlihy2020art}
traverses a linked list while holding at most two locks: it acquires the lock on
the successor before releasing the lock on the predecessor. Holding a lock on
\texttt{curr} permits safe access to \texttt{curr->next} for key comparisons or
boundary checks. During inserts, both locks are held to link a new node between
\texttt{curr} and \texttt{curr->next}. Lock coupling extends naturally to
\skiplists, which are towers of linked lists.

\para{Reader-writer (RW) locks} Reader-writer locks (RW
locks)~\cite{courtois1971concurrent} allow concurrent shared access via
\texttt{read\_lock()} or exclusive access via \texttt{write\_lock()}. Multiple
threads may hold the lock in read mode simultaneously; any thread holding it in
write mode blocks all others. Algorithm~\ref{alg:hoh} shows one HOH traversal
step under RW locks.

\para{RW-lock-based concurrency control in the \bskiplist} The RW-lock-based
\bskiplist~\cite{bskip} uses lock coupling top-to-bottom and left-to-right,
following the top-down insertion algorithm. It exploits upfront knowledge of the
promotion level to take write locks 
starting at the promotion level.

\para{OLC} An optimistic lock packs a write bit and a
version number into a single 64-bit word~\cite{leis2016art,
  leis2019optimistic}.
The LSB acts as the write bit; writers set it on lock acquisition and increment
the version (clearing the LSB) on release.

In OLC, readers detect concurrent
modifications by comparing the version before and after accessing a node without
modifying shared state, eliminating cache-line invalidations. Readers may
occasionally retry, but retries are rare in practice.
OLC
is a form of \defn{optimistic concurrency control} (OCC)~\cite{KungRo81}:
threads traverse without locks and restart if they detect a concurrent
modification before committing.

\begin{algorithm}[t]
\caption{Optimistic lock coupling step}
\label{alg:olc}
\begin{algorithmic}[1]
\State \textbf{Input:} $x$: current node, $v_x$: version of $x$ read in previous
iteration
\State $y \gets x.next$
\State $v_y \gets \LOADVERSION{y}$
\If{$\IsWriteLocked{v_x} \lor v_x \neq \LOADVERSION{x}$}
  \State {\textbf{goto} retry}
\EndIf
    \State $x \gets y$
    \State $v_x \gets v_y$
\end{algorithmic}
\end{algorithm}

\para{OLC traversals} OLC traversals follow the same order as HOH but replace
read locking with version validation~\cite{leis2019optimistic}. A thread reads
the next node's version (analogous to acquiring the next lock), then validates
the current node by checking that its version is unchanged and not write-locked
(analogous to releasing the current lock). Algorithm~\ref{alg:olc} shows one
traversal step; validation failure triggers a retry from the root.


\section{Proposed node structure.}\label{sec:bskip-insert}

This section contrasts the node layouts of the RW-lock-based
\bskiplist~\cite{bskip} and the proposed optimistic \bskiplist.
We also address practical concerns such as version overflow and the ABA problem.

\begin{figure}[t]
    \centering
    \includegraphics[width=\columnwidth]{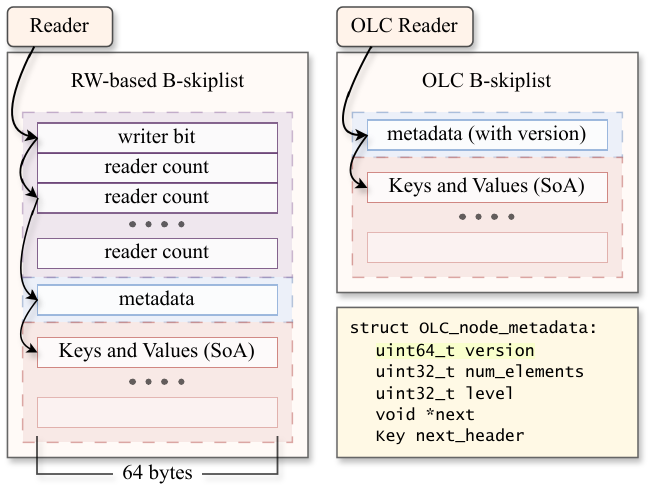}
    \caption{Node layouts in the \rwbased \bskiplist and the
      proposed \obskipnospace. }
    \label{fig:node_layout_comparison}
  \end{figure}

In the \rwbased \bskiplist, each node begins with the partitioned
RW lock, followed by metadata (number of elements, node level, next pointer, and
first key of the next node), then keys and values in structure-of-arrays (SoA)
layout. Reading a node requires at least two cache-line accesses for the write bit and reader counter before reaching
any useful data.

In the proposed \obskipnospace, each node begins with the 8-byte version number,
followed by metadata and SoA-layout keys and values. The first cache-line access
thus contains the version number and complete metadata, reducing both memory
latency and footprint compared to the RW-lock-based design.

\para{Version overflow and ABA} The \defn{ABA problem} arises when a version
number wraps around to a previously-seen value, causing a reader to incorrectly
conclude that no modification occurred. With a 64-bit version, this is not a
practical concern: at $10^9$ modifications per second, a full wraparound takes
$2^{63}/10^9$ seconds $\approx$ 292 years. Even if
wraparound did occur, the version continues from zero, which is safe under the
same argument.


\section{Optimistic algorithms.}\label{sec:concurrency}


This section presents the algorithms and design of the \obskip
(see~\secref{prelim}), detailing the Find and Insert algorithms and how they
maintain consistency under OLC. We also discuss deadlock freedom and the memory
reclamation mechanism required to support lock-free readers.

OLC is effective for the \bskiplist because most writes affect only the
leaves: the number of elements in internal nodes decreases exponentially
by level, so write conflicts at upper
levels are exceedingly rare. Even when a retry occurs, the traversal path is
likely still in cache, keeping the retry cost low.

\subsection{Algorithms.}

\para{Find} Algorithm~\ref{alg:find} shows the \textsc{Find} operation. The
algorithm begins at the root, loading its version before entering the
traversal loop; $\VALIDATE{x}{v}$ issues a \textsc{LoadVersion} on node $x$ and
checks that the result equals $v$. At each level, the algorithm moves right
while the next header does not exceed the search key, following the OLC step in
Algorithm~\ref{alg:olc}.

At internal levels ($i>0$), descent uses a two-phase validation. First, the
algorithm validates the current node before dereferencing the child pointer
(line \ref{line:algo_find_phase1}). This is necessary because a concurrent
writer may leave the pointer in an indeterminate state, making an unvalidated
dereference unsafe. Second, the algorithm validates the current node again after
loading the child's version, completing the OLC handoff (line
\ref{line:algo_find_phase2}). At the leaf level ($i=0$), the algorithm searches
for the key locally, validates the node before returning, and returns the
associated value if found.

\begin{algorithm}[!t]
\caption{$\text{Find}(L, k)$}
\label{alg:find}
\begin{algorithmic}[1]
\Require Skiplist $L$, key $k$
\Ensure Value associated with $k$, or \textsc{Nil}
\State \textbf{retry:}
\State $x \gets L.root$
\State $v_x \gets \textsc{LoadVersion}(x)$
\For{$i \gets MAX\_LEVEL - 1$ \textbf{to} $0$}
    \While{$x.next\_header \leq k$}
        \State $y \gets x.next$
        \If{$y = \textsc{Nil}$}
            \State \textbf{goto} retry
        \EndIf
        \State $v_y \gets \textsc{LoadVersion}(y)$
        \If{$\IsWriteLocked{v_x} \lor \lnot\textsc{Validate}(x, v_x)$}
            \State \textbf{goto} retry
        \EndIf
        \State $x \gets y$, $v_x \gets v_y$
    \EndWhile
    \If{$\textsc{IsWriteLocked}(v_x)$}
        \State \textbf{goto} retry
    \EndIf
    \State $r \gets \textsc{FindRank}(x.keys, k)$
    \If{$i = 0$}
        \If{$x.keys[r] = k$}
            \State $val \gets x.values[r]$
            \If{$\lnot\textsc{Validate}(x, v_x)$}
                \State \textbf{goto} retry
            \EndIf
            \State \Return $val$
        \Else
            \If{$\lnot\textsc{Validate}(x, v_x)$}
                \State \textbf{goto} retry
            \EndIf
            \State \Return \textsc{Nil}
        \EndIf
    \Else
        \State $z \gets x.children[r]$
        \If{$z = \textsc{Nil}$}
            \State \textbf{goto} retry
        \EndIf
        \LineComment{Phase 1: validate child pointer}
        \If{$\lnot\textsc{Validate}(x, v_x)$} \label{line:algo_find_phase1}
            \State \textbf{goto} retry
        \EndIf
        \State $v_z \gets \textsc{LoadVersion}(z)$
        \LineComment{Phase 2: validate child version}
        \If{$\lnot\textsc{Validate}(x, v_x)$} \label{line:algo_find_phase2}
            \State \textbf{goto} retry
        \EndIf
        \State $x \gets z$, $v_x \gets v_z$
    \EndIf
\EndFor
\State \Return \textsc{Nil}
\end{algorithmic}
\end{algorithm}

\para{Range query} The \textsc{Scan} operation extends \textsc{Find}. It takes
in a starting key $k$, an accumulator, a callback function (e.g., \textsc{Sum}),
and a stop condition such as a max length or an end key. The algorithm first
performs a \textsc{Find} to locate the leaf node \texttt{curr} containing the
largest key $\leq k$. It then iteratively applies the callback to each key-value
pair until the stop condition is met or the node ends. Upon reaching the end of
the leaf node, the algorithm validates \texttt{curr}, reads \texttt{curr.next},
and continues until the stop condition is met or the last leaf node is
reached. If any validation fails, the accumulator is reset and the operation
retries from the root.

\begin{figure}[t]
    \centering
    \includegraphics[width=\columnwidth]{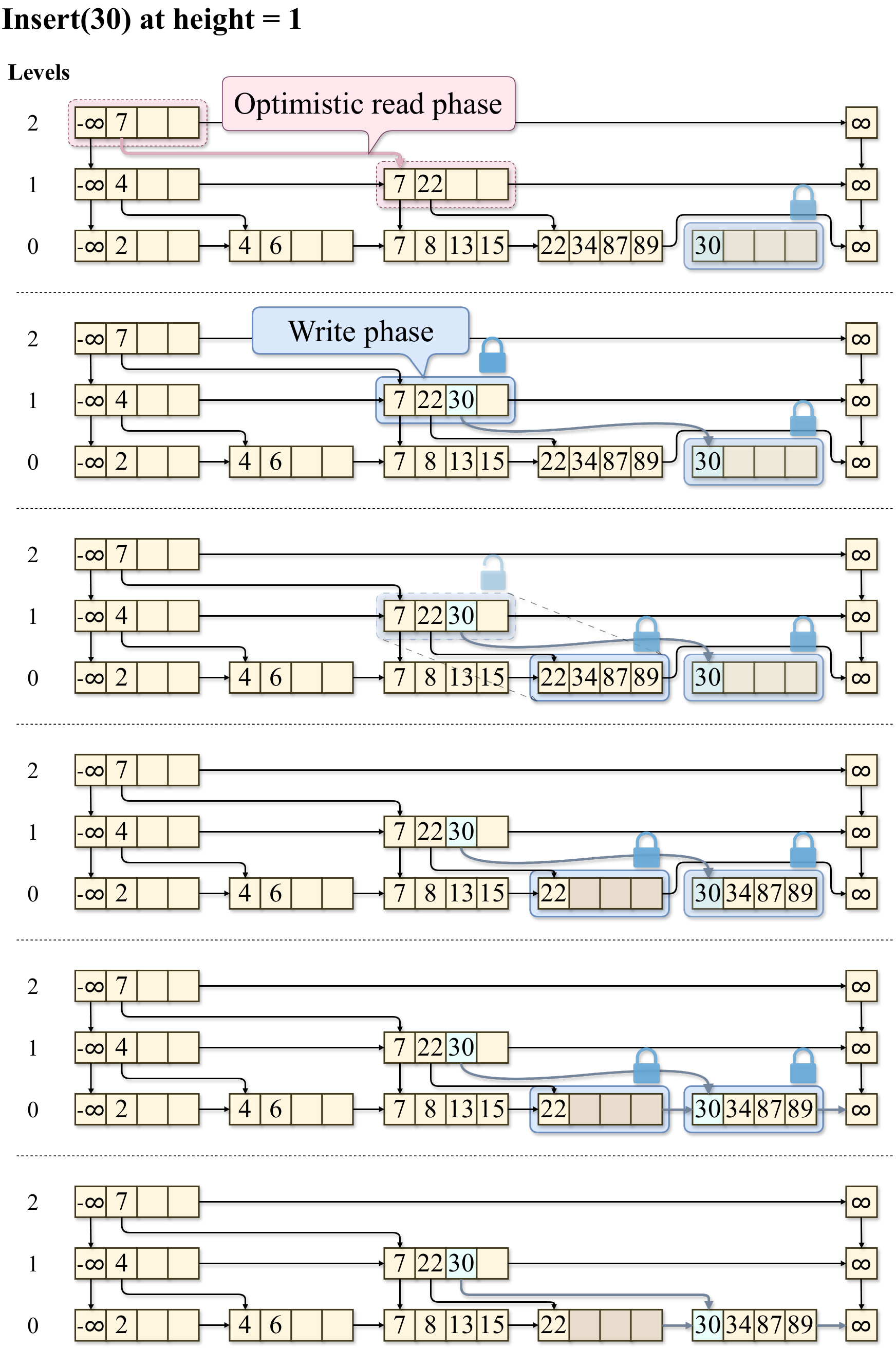}
    \caption{An example of Insert(30) at height = 1. The hybrid OLC protocol consists of two phases. The \emph{read} phase traverses optimistically from the top level down to the promotion level. Upon reaching the promotion level, the
    \emph{write} phase switches to exclusive locks.}
    \label{fig:insert}
\end{figure}

\begin{algorithm}[!t]
\caption{$\text{Insert}(L \text{: Skiplist}, k \text{: Key}, v \text{: Value})$}
\label{alg:insert}
\begin{algorithmic}[1]
\State $h \gets \textsc{PromotionLevel}(k)$
\Statex $new\_nodes \gets$ allocate $h$ write-locked nodes
\LineComment After optimistic traversal, $x$ is the node at promotion level $h$ holding a \textbf{write lock}
\State $r \gets \textsc{FindRank}(x.keys, k)$
\If{$h = 0$}
    \State $\textsc{InsertLeaf}(x, r+1, k, v)$ \label{line:algo_insert_plevel=0}
    \State $\textsc{WriteUnlock}(x)$
    \State \Return
\Else
    \State $\textsc{InsertInternal}(x, r+1, k, new\_nodes[h-1])$
    \State $parent \gets x$, $x \gets x.children[r]$
\EndIf
\While{$h > 0$} \label{line:algo_insert_promotion_start}
    \Comment{Top-down promotion}
    \State $h \gets h - 1$
    \State $\textsc{WriteLock}(x)$
    \If{$parent \neq \textsc{Nil}$}
        \State $\textsc{WriteUnlock}(parent)$
        \State $parent \gets \textsc{Nil}$
    \EndIf
    \While{$x.next\_header \leq k$} \label{line:algo_insert_horiz_start}
        \State $\textsc{WriteLock}(x.next)$
        \State $x' \gets x$, $x \gets x.next$
        \State $\textsc{WriteUnlock}(x')$
    \EndWhile \label{line:algo_insert_horiz_end}
    \State $r \gets \textsc{FindRank}(x.keys, k)$
    \State $p \gets new\_nodes[h]$ \label{line:algo_insert_node_p}
    \State $p.next\_header \gets x.next\_header$
    \State $p.next \gets x.next$, $p.level \gets h$
    \LineComment{Insert $k$ into $p$ before reading $p.keys[0]$}
    \If{$h > 0$}
        \State $\textsc{InsertInternal}(p, 0, k, new\_nodes[h-1])$
    \Else
        \State $\textsc{InsertLeaf}(p, 0, k, v)$
    \EndIf
    \LineComment{Move $x.keys[r+1..]$ into $p$ at slot 1}
    \State $\textsc{SplitKeys}(x, p, r+1)$
    \State $x.next\_header \gets p.keys[0]$ \Comment{$p.keys[0]=k$}
    \State $x.next \gets p$ \label{line:algo_insert_link}
    \If{$h > 0$}
        \State $parent \gets x$, $x \gets x.children[r]$
    \Else
        \State $\textsc{WriteUnlock}(x)$
    \EndIf
    \State $\textsc{WriteUnlock}(p)$ \label{line:algo_insert_promotion_end}
\EndWhile
\State \Return
\end{algorithmic}
\end{algorithm}


\para{Insertion} Algorithm~\ref{alg:insert} shows the \emph{hybrid} OLC-based
Insert operation. Writers begin with an optimistic \emph{read} phase, using the
same traversal as \textsc{Find}, until their promotion level. The key insight is
that the promotion height is known \emph{upfront} due to the \bskiplist's
probabilistic nature.

Since promotions occur with geometrically decreasing probability, most
insertions reach low heights, so the optimistic traversal covers the vast
majority of levels, minimizing cache invalidations and write-lock hold time.

Upon reaching the promotion level, the algorithm acquires a write lock on node
$x$ and transitions to the \emph{write} phase.
If the promotion level is zero (i.e., the leaf level), the key-value pair is
inserted directly (line~\ref{line:algo_insert_plevel=0}) and the operation
terminates. Otherwise, the algorithm enters a top-down promotion phase
(lines~\ref{line:algo_insert_promotion_start}--\ref{line:algo_insert_promotion_end}),
processing one level at a time while $h > 0$.

At each level, the algorithm locates the correct position by traversing
horizontally under HOH locking
(lines~\ref{line:algo_insert_horiz_start}--\ref{line:algo_insert_horiz_end}),
then splits $x$ and links in a new node $p$
(lines~\ref{line:algo_insert_node_p}--\ref{line:algo_insert_link}). Since the
promotion height is known upfront, the correct number of new nodes is
preallocated and write-locked before traversal. The insertion then proceeds level-by-level to the leaves.

Figure~\ref{fig:insert} illustrates the hybrid CC protocol. The optimistic read
phase (top panel) descends through upper levels by checking version numbers
without acquiring write locks. Upon reaching the promotion level, the algorithm
acquires a write lock on the target node and transitions to the write phase
(second panel). The subsequent panels show how locks are acquired
level-by-level as the new node is linked into the
structure.

\para{Deletion} Like \textsc{Insert}, \textsc{Delete}($k$) begins with an optimistic read phase, descending from the top level to $k$'s promotion level. At that level, the algorithm acquires a write lock and switches to the write phase, using HOH locking to locate the node containing $k$, read its child $c$, and delete $k$ from it. A \bskip invariant guarantees that $c$ also contains $k$, so the algorithm repeats this process level-by-level down to the leaf. After each deletion, if the node is now empty, the algorithm unlinks it by updating its predecessor's \texttt{next} pointer.

Unlike the \rwbskip, the \olcbskip cannot immediately deallocate a disconnected node, because optimistic readers may still hold references to it. To support safe memory reuse, the \olcbskip uses epoch-based reclamation (EBR)~\cite{fraser2004practical}. EBR defers reclamation until all threads that could have observed a node have passed through a safe point, ensuring readers always access valid memory.

\subsection{Correctness.}

\para{Skiplist structure} Following prior work~\cite{bskip}, we relax the \defn{inclusion invariant} ---
the property that every element at level $h$ must appear at all lower levels ---
only \emph{during} insertions. Under the top-down algorithm, an element may appear at higher levels before
lower ones. The invariant is restored upon completion of any insertion, and the final
pointer structure matches that of a bottom-up algorithm.

This relaxation does not affect the theoretical bounds, which depend on the expected height and number of nodes traversed per level and are determined by the probabilistic coin flips that fix each
element's promotion level, not the insertion order.

\para{Deadlock freedom} Since readers operate entirely under OLC
without acquiring locks, they cannot participate in a deadlock. Writers use a spinlock with the same top-down HOH algorithm as the \rwbased
\bskiplist; because locks are acquired strictly top-down, no cycle can form
among waiting writers.
Correctness follows from the coupling discipline: every move to the next node
is performed under a valid coupling --- holding the current lock (HOH) or
validating the current version (OLC) --- ensuring the traversal path remains
consistent at every step.


\section{Evaluation.}\label{sec:eval}
This section evaluates the \olcbskip on YCSB~\cite{ycsb}, TPC-C, and delete
workloads against the \rwbased
\bskiplist~\cite{bskip}, an optimistic \skiplist from Facebook's \folly
library~\cite{folly}, an \rwbased optimistic \bplustree~\cite{XuLiWh23}, and an
OLC-based \bplustree~\cite{leis2019optimistic, WangPaLi18}.

Overall, the \olcbskip achieves \tpAll the throughput of the other tested
indices, up to \latMax lower latency than the \skiplists and the \rwbtree, and
latency within \latHigher of the \olcbtree.
Furthermore, the \olcbskip achieves \olcbskipscalingA, \olcbskipscalingC, and
\olcbskipscalingL self-relative speedup from 1 to 128 threads on
YCSB~\cite{ycsb-core-workloads} workloads A, C, and the load phase,
respectively.

\subsection{Setup.}

\para{Hardware} We ran the experiments on a server with an AMD EPYC 9555 64-Core
Processor with 1007 GB of memory. The server has a 3 MiB L1 data cache, a 2 MiB
L1 instruction cache, a 64 MiB L2 cache, and a 256 MiB L3 cache. The 64 cores
are distributed on 4 NUMA nodes and support a maximum of 128 threads. To
minimize NUMA effects, we ran 1--16-thread experiments on a single NUMA node,
32-thread experiments on two NUMA nodes, and 64- and 128-thread experiments on
all four NUMA nodes.

All times are the median of 5 trials after one warm-up trial. To measure
latency, each thread measures the average time taken for a batch of 10
operations\footnote{We measure the average of 10 operations instead of
  individual operations to preserve the contention between threads.} and stores
it in a thread-safe vector. After all operations complete, we sort the
per-thread measurements and compute latency at each percentile.

\begin{table}[t]
  \centering \resizebox{.8\columnwidth}{!}{
\begin{tabular}{@{}ll@{}}
 \hline
\textit{Workload} & \textit{Description} \\
 \hline
Load & 100\% inserts from empty\\
A&50\% finds, 50\% inserts \\
B&95\% finds, 5\% inserts \\
C&100\% finds  \\
E&\makecell{95\% short range iterations\\ (\maxlen = 100), 5\% inserts} \\
 \hline
\end{tabular}
}
\caption{YCSB workload descriptions.}
\label{tab:ycsb-workloads}
\end{table}

\para{Workloads} We follow the standard YCSB~\cite{ycsb-core-workloads}
procedure, which consists of a load phase and a run phase. In the load phase,
100 million (100M) elements are inserted into an empty index. In the run phase,
a mixture of 100M operations is executed, with the mixture determined by the
YCSB workload. Operations within each phase run concurrently; the phases run
sequentially.

\tabref{ycsb-workloads} presents details of the selected YCSB workloads. We
tested workloads\footnote{We omit workload D from YCSB because it benchmarks the
  read-latest operation, which is not the focus of this work.} A, B, C, and E
from the core YCSB workloads generated with RECIPE~\cite{LeeEtAl19}.

We evaluate each workload under both uniform and zipfian distributions in the run
phase. In the uniform workloads, elements in both phases are drawn from a uniform
distribution. In the zipfian workloads, load-phase elements are uniform while
run-phase \emph{query} elements are drawn from a zipfian distribution.

Additionally, we tested all
indices on the Transaction Processing Performance Council's TPC-C
benchmark~\cite{leutenegger1993modeling}, which simulates large-scale online
business activity. Specifically, we used the
implementation\footnote{\url{https://github.com/tomercory/DBx1000_foresight}}~\cite{cory2026skiplists}
from the DBx1000 in-memory database~\cite{yu2014staring}. It supports a
representative sample of the TPC-C benchmark with only Payment and New Order
transactions (which account for 88\% of the transactions in TPC-C). Our
evaluation was conducted using 128 warehouses and 128 threads. Execution
terminates once any thread reaches 100,000 transactions, yielding a total of
8 to 11 million completed transactions across the different indices.

The YCSB and TPC-C benchmarks do not include deletion, so we generated two
delete workloads that mirror the same write percentages as YCSB workloads A
(50\%/50\% finds/deletes) and B (95\%/5\% finds/deletes), i.e., with deletes
instead of inserts. To ensure that the keys to be deleted are in the index, we used the
keys from the run phase of YCSB workload C (100\% finds).

\para{Systems setup} To ensure fairness, all test drivers execute concurrent operations
using pthreads~\cite{nichols1996pthreads}, and all keys and values are 64-bit
integers.

We use the \rwbased \bskiplist\footnote{
  \url{https://github.com/Ratbuyer/bskip_artifact}} (\rwbskip,
\defn{\rwbskipplot}) in \texttt{C++} as our primary baseline~\cite{bskip}, with
default parameters\footnote{The default parameters set node size = 2048 bytes,
  promotion probability factor = 0.5, and max height = 5.}.

We implemented the \olcbskip\footnote{\url{https://github.com/illliliiill1116/hocc64-bskiplist/tree/artifact-eval}} (\defn{\obskipplot}) in \texttt{C} using the same
parameters as the \rwbskip~\cite{bskip}. To isolate performance differences to
the CC scheme, we adopted the \rwbskip's fixed-size nodes but omit that
description from~\secref{concurrency} for simplicity. Fixed-size nodes require a
slight modification of the insert algorithm and CC scheme, but do not affect the
key feature that writes only occur up to the promotion level.

We use a \texttt{C++} concurrent skiplist\footnote{
  \url{https://github.com/facebook/folly/blob/main/folly}} from Facebook's
\folly library (\defn{\follyplot}) as a representative high-performance
non-blocked (one key per node) \skiplist~\cite{folly}. Folly's \skiplist
uses optimistic concurrency control: reads are lock-free and mostly wait-free,
while writes require locking~\cite{herlihy2006provably, herlihy2007simple}.
We use the native interface for point operations and the iterator interface for
range queries (since it lacks native range-query support), with initial height
set to 20. Folly's skiplist only stores keys.

We also tested a \texttt{C++} concurrent
\bplustree\footnote{\url{https://github.com/wheatman/BP-Tree/tree/main/tlx-plain/container}}~\cite{XuLiWh23}
based on RW locks and OCC~\cite{KungRo81} (\rwbtree, \defn{\rwbtreeplot}). Its
OCC protocol uses HOH read locks to traverse from the root to the second-to-last
level, then acquires write locks at the leaf to insert the key. If a split is
required, the algorithm relinquishes all locks and retries from the root under
write locks.

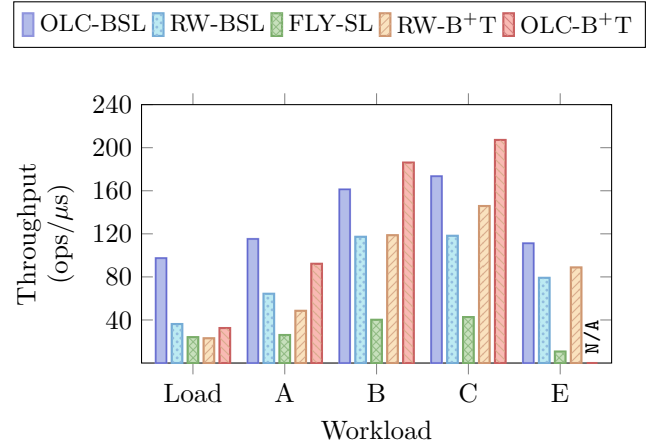
\begin{figure}[t]
  \centering
  \begin{tikzpicture}
    \begin{axis}[
    width=7.8cm, height=5cm,
        ybar,
        legend style={at={(0.4, 1.4)}, anchor=north},
        ylabel={Throughput (ops/$\mu$s)},
         ylabel style={align=center, text width=3cm},
        xlabel={Workload},
        xtick=data,
        bar width=4pt,
        legend columns=5,
        legend style={font=\small},
        ytick={40, 80, 120, 160, 200, 240},
        enlarge x limits=0.14,
        symbolic x coords={Load, A, B, C, E},
        x tick label style={text width = 1cm, align = center},
        ymin = 0,
        ymax = 240,
      ]

    \addplot [fill=o-bskip, draw=o-bskip-dark!70, thick, forget plot]
    coordinates {
        (Load, 97.40)(A, 115.23)(B, 161.25)(C, 173.49)(E, 111.2)
    };
    \addplot [
        fill=o-bskip,
        draw=o-bskip-dark,
        pattern=none,
        pattern color=o-bskip-dark!70,
        legend image code/.code={
            \draw[fill=o-bskip, draw=o-bskip-dark!70, thick] (-0.07cm,-0.15cm) rectangle (0.07cm,0.15cm);
            \draw[pattern=none, pattern color=o-bskip-dark!70, draw=none] (-0.07cm,-0.15cm) rectangle (0.07cm,0.15cm);
        }
    ]
    coordinates {
        (Load, 97.40)(A, 115.23)(B, 161.25)(C, 173.49)(E, 111.2)
    };
    \addlegendentry{\obskipplot}

    \addplot [fill=bskip, draw=bskip-dark, thick, forget plot]
    coordinates {
        (Load, 36.23)(A, 64.36)(B, 117.27)(C, 118.21)(E, 79.16)
    };
    \addplot [
        fill=bskip,
        draw=bskip-dark,
        pattern=crosshatch dots,
        pattern color=bskip-dark!80,
        legend image code/.code={
            \draw[fill=bskip, draw=bskip-dark, thick] (-0.07cm,-0.15cm) rectangle (0.07cm,0.15cm);
            \draw[pattern=crosshatch dots, pattern color=bskip-dark!80, draw=none] (-0.07cm,-0.15cm) rectangle (0.07cm,0.15cm);
        }
    ]
    coordinates {
        (Load, 36.23)(A, 64.36)(B, 117.27)(C, 118.21)(E, 79.16)
    };
    \addlegendentry{\rwbskipplot}

    \addplot [fill=folly, draw=folly-dark, thick, forget plot]
    coordinates {
         (Load, 24.03)(A, 26.03)(B, 40.19)(C, 42.68)(E, 10.69)
    };
    \addplot [
        fill=folly,
        draw=folly-dark,
        pattern=crosshatch,
        pattern color=folly-dark!80,
        legend image code/.code={
            \draw[fill=folly, draw=folly-dark, thick] (-0.07cm,-0.15cm) rectangle (0.07cm,0.15cm);
            \draw[pattern=crosshatch, pattern color=folly-dark!80, draw=none] (-0.07cm,-0.15cm) rectangle (0.07cm,0.15cm);
        }
    ]
    coordinates {
         (Load, 24.03)(A, 26.03)(B, 40.19)(C, 42.68)(E, 10.69)
    };
    \addlegendentry{\follyplot}

    \addplot [fill=btree, draw=btree-dark, thick, forget plot]
    coordinates {
        (Load, 22.99)(A, 48.53)(B, 118.72)(C, 145.84)(E, 88.80)
    };
    \addplot [
        fill=btree,
        draw=btree-dark,
        pattern=north east lines,
        pattern color=btree-dark!80,
        legend image code/.code={
            \draw[fill=btree, draw=btree-dark, thick] (-0.07cm,-0.15cm) rectangle (0.07cm,0.15cm);
            \draw[pattern=north east lines, pattern color=btree-dark!80, draw=none] (-0.07cm,-0.15cm) rectangle (0.07cm,0.15cm);
        }
    ]
    coordinates {
        (Load, 22.99)(A, 48.53)(B, 118.72)(C, 145.84)(E, 88.80)
    };
    \addlegendentry{\rwbtreeplot}


    \addplot [fill=olc-btree, draw=olc-btree-dark, thick, forget plot]
    coordinates {
        (Load, 32.5)(A, 92.2)(B, 186.2)(C, 207.2)(E, 0)
    };
    \addplot [
        fill=olc-btree,
        draw=olc-btree-dark,
        pattern=north west lines,
        pattern color=btree-dark!80,
        nodes near coords,
        nodes near coords style={
            rotate=90,
            anchor=west,
            font=\footnotesize,
        },
        point meta=explicit symbolic,
        legend image code/.code={
            \draw[fill=olc-btree, draw=olc-btree-dark, thick] (-0.07cm,-0.15cm) rectangle (0.07cm,0.15cm);
            \draw[pattern=north west lines, pattern color=olc-btree-dark!80, draw=none] (-0.07cm,-0.15cm) rectangle (0.07cm,0.15cm);
        }
    ]
    coordinates {
        (Load, 32.5)(A, 92.2)(B, 186.2)(C, 207.2)(E, 0)[\texttt{N/A}]
    };
    \addlegendentry{\olcbtreeplot}


    \end{axis}
  \end{tikzpicture}
  \vspace{-.5cm}
  \caption{Throughput on zipfian YCSB workloads on 128 threads.}
    \label{fig:tp_zipfian}
  \end{figure}

Finally, we tested a \texttt{C++} OLC-based
\bplustree\footnote{\url{https://github.com/wangziqi2016/index-microbench/tree/master/BTreeOLC}}
(\olcbtree, \defn{\olcbtreeplot})~\cite{leis2019optimistic, WangPaLi18}.  The
implementation does not support multi-node range queries, so we omit workload E
for this index.

We compiled all baselines using g++ 13.3.0 and the \olcbskip using gcc 13.3.0
(all with \texttt{-Ofast}).

\subsection{Throughput.}
\begin{figure*}[t]
  \centering

  \begin{tikzpicture}
  \begin{axis}[
    hide axis,
    xmin=0, xmax=1,
    ymin=0, ymax=1,
    legend columns=5,
    legend style={
      draw=black,
      fill=white,
      inner sep=2pt,
      /tikz/every even column/.append style={column sep=0.4cm}
    }
  ]

  \addlegendimage{mark=square*, color=o-bskip-dark, thick}
  \addlegendentry{\obskipplot}

  \addlegendimage{mark=triangle*, color=bskip-dark, thick}
  \addlegendentry{\rwbskipplot}

  \addlegendimage{mark=*, color=folly-dark, thick}
  \addlegendentry{\follyplot}

  \addlegendimage{mark=diamond*, color=btree-dark, thick}
  \addlegendentry{\rwbtreeplot}

  \addlegendimage{mark=pentagon*, color=olc-btree-dark, thick}
  \addlegendentry{\olcbtreeplot}

  \end{axis}
  \end{tikzpicture}


  \begin{subfigure}{0.48\textwidth}
    \centering
    \begin{tikzpicture}
    \begin{axis}[
      width=0.97\linewidth, height=4cm,
      axis lines = left,
      xlabel = Percentile, ylabel = {Latency ($\mu$s)},
      ymode=log, log ticks with fixed point,
      ymin=0, ymax=15,
      ytick={2,4,10,15},
      xtick distance=1,
      xticklabels={0,50,90,99,99.9}, xmax=4
    ]

    \addplot [mark=square*, color=o-bskip-dark, thick]
    coordinates {(1,1.292)(2,1.488)(3,1.752)(4,2.282)};

    \addplot [mark=triangle*, color=bskip-dark, thick]
    coordinates {(1,1.198)(2,1.399)(3,1.951)(4,3.340)};

    \addplot [mark=*, color=folly-dark, thick]
    coordinates {(1,4.468)(2,5.110)(3,5.779)(4,6.342)};

    \addplot [mark=diamond*, color=btree-dark, thick]
    coordinates {(1,1.229)(2,1.528)(3,5.572)(4,13.688)};

    \addplot [mark=pentagon*, color=olc-btree-dark, thick]
    coordinates {(1,0.847)(2,0.953)(3,1.426)(4,1.958)};

    \end{axis}
    \end{tikzpicture}
    \caption{Uniform workload A (50\%/50\% finds/inserts)}
  \end{subfigure}
  \hfill
  \begin{subfigure}{0.48\textwidth}
    \centering
    \begin{tikzpicture}
    \begin{axis}[
      width=0.97\linewidth, height=4cm,
      axis lines = left,
      xlabel = Percentile, ylabel = {Latency ($\mu$s)},
      ymode=log, log ticks with fixed point,
      ymin=0, ymax=15,
      ytick={2,4,10,15},
      xtick distance=1,
      xticklabels={0,50,90,99,99.9}, xmax=4
    ]

    \addplot [mark=square*, color=o-bskip-dark, thick]
    coordinates {(1,0.992)(2,1.124)(3,1.324)(4,1.645)};

    \addplot [mark=triangle*,color=bskip-dark, thick]
    coordinates {(1,0.960)(2,1.076)(3,1.287)(4,2.038)};

    \addplot [mark=*, color=folly-dark, thick]
    coordinates {(1,3.424)(2,3.894)(3,4.455)(4,4.972)};

    \addplot [mark=diamond*, color=btree-dark, thick]
    coordinates {(1,0.894)(2,0.982)(3,1.112)(4,2.086)};

    \addplot [mark=pentagon*, color=olc-btree-dark, thick]
    coordinates {(1,0.665)(2,0.732)(3,0.816)(4,1.618)};

    \end{axis}
    \end{tikzpicture}
    \caption{Uniform workload C (100\% finds)}
  \end{subfigure}
  \vspace{0.3cm}

  \begin{subfigure}{0.48\textwidth}
    \centering
    \begin{tikzpicture}
    \begin{axis}[
      width=0.97\linewidth, height=4cm,
      axis lines = left,
      xlabel = Percentile, ylabel = {Latency ($\mu$s)},
      ymode=log, log ticks with fixed point,
      ymin=0, ymax=15,
      ytick={2,4,10,15},
      xtick distance=1,
      xticklabels={0,50,90,99,99.9}, xmax=4
    ]

    \addplot [mark=square*, color=o-bskip-dark, thick]
    coordinates {(1,1.147)(2,1.391)(3,1.654)(4,2.174)};

    \addplot [mark=triangle*, color=bskip-dark, thick]
    coordinates {(1,1.113)(2,1.350)(3,1.907)(4,3.828)};

    \addplot [mark=*, color=folly-dark, thick]
    coordinates {(1,4.107)(2,4.864)(3,5.636)(4,6.302)};

    \addplot [mark=diamond*, color=btree-dark, thick]
    coordinates {(1,1.089)(2,1.388)(3,5.012)(4,13.038)};

    \addplot [mark=pentagon*, color=olc-btree-dark, thick]
    coordinates {(1,0.744)(2,0.882)(3,1.320)(4,1.857)};

    \end{axis}
    \end{tikzpicture}
    \caption{Zipfian workload A (50\%/50\% finds/inserts)}
  \end{subfigure}
  \hfill
  \begin{subfigure}{0.48\textwidth}
    \centering
    \begin{tikzpicture}
    \begin{axis}[
      width=0.97\linewidth, height=4cm,
      axis lines = left,
      xlabel = Percentile, ylabel = {Latency ($\mu$s)},
      ymode=log, log ticks with fixed point,
      ymin=0, ymax=15,
      ytick={2,4,10,15},
      xtick distance=1,
      xticklabels={0,50,90,99,99.9}, xmax=4
    ]

    \addplot [mark=square*, color=o-bskip-dark, thick]
    coordinates {(1,0.747)(2,0.925)(3,1.106)(4,1.409)};

    \addplot [mark=triangle*, color=bskip-dark, thick]
    coordinates {(1,0.788)(2,0.940)(3,1.124)(4,1.813)};

    \addplot [mark=*, color=folly-dark, thick]
    coordinates {(1,2.822)(2,3.427)(3,4.020)(4,4.582)};

    \addplot [mark=diamond*, color=btree-dark, thick]
    coordinates {(1,0.711)(2,0.837)(3,0.970)(4,1.797)};

    \addplot [mark=pentagon*, color=olc-btree-dark, thick]
    coordinates {(1,0.500)(2,0.609)(3,0.713)(4,1.467)};

    \end{axis}
    \end{tikzpicture}
    \caption{Zipfian workload C (100\% finds)}
  \end{subfigure}
  \vspace{-0.3cm}
  \caption{Latency at different percentiles on uniform and zipfian YCSB workloads on 128 threads.} 
  \label{fig:combined_latencies}

\end{figure*}
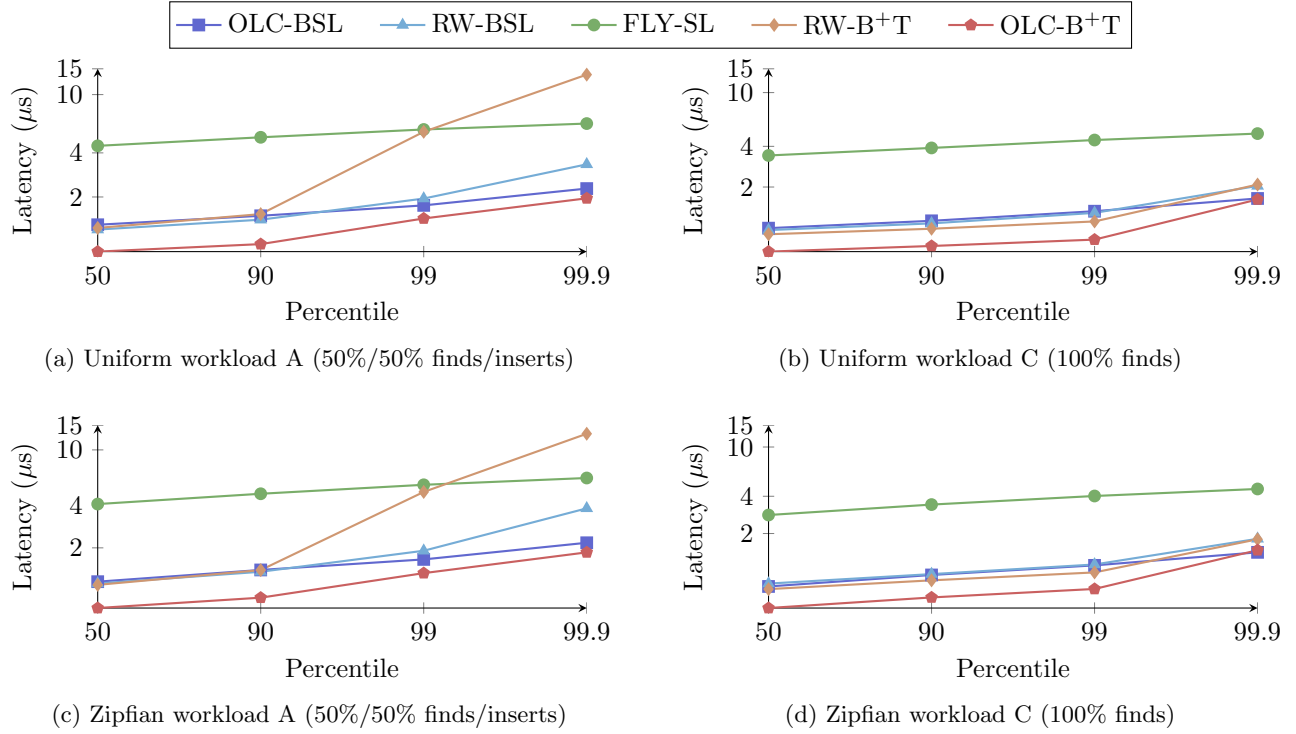

\para{YCSB benchmarks}
Figures \ref{fig:tp_uniform} and \ref{fig:tp_zipfian} report throughput on
uniform and zipfian YCSB workloads. Overall, the \olcbskip achieves speedups over
\skiplist-based indices and the \rwbtree\xspace on all workloads. Furthermore, it is
faster than the \olcbtree\xspace on write-heavy workloads (load and workload A), and
slightly slower on read-heavy workloads (workloads B and C). All systems achieve
higher throughput on zipfian workloads than on uniform workloads because query
paths have better temporal locality under skewed distributions.

For the load phase, the \olcbskip is about \tpBskipL, \tpFollyL, \tpRWBTreeL,
and \tpOLCBTreeL faster than the \rwbskip, \folly, \rwbtree, and \olcbtree,
respectively. On point-operation workloads (A, B, and C), the \olcbskip is
about \tpBskipABC faster than the \rwbskip, \tpFollyABC faster than \folly,
\tpRWBTreeABC faster than the \rwbtree, and achieves \tpOLCBTreeABC the
throughput of the \olcbtree. On insert/range-query workload E, the \olcbskip is around \tpBskipE
faster than the \rwbskip, \tpFollyE faster than \folly, and \tpRWBTreeE faster
than the \rwbtree.

\para{YCSB discussion} The \olcbskip improves cache efficiency over the \rwbskip by
fitting all concurrency metadata (version number and write bit) on a single
cache line. \tabref{intro_table} shows that on the load phase and workload A, the \olcbskip incurs
$1.1\times$ fewer cache misses and $1.8\times$ fewer cache references than the
\rwbskip.

In practice, retries are rare, enabling the \olcbskip to scale well on
write-heavy workloads. During the load phase (100\% inserts, the worst case for
optimism), at most $6\%$ of operations retry. In workload A (50\%/50\%
finds/inserts), about $1\%$ retry. In workloads B and E (95\%/5\% reads/inserts),
retries drop to about $0.5\%$.

The \olcbskip is especially well-suited to write-heavy workloads: its hybrid
scheme acquires write locks only from the promotion level down, exploiting
upfront knowledge of promotion height to minimize lock hold time. The \olcbskip
achieves $1.2$--$4.3\times$ higher throughput on write-heavy workloads (load and workload
A) compared to the \bplustree-based indices, which must lock multiple nodes
simultaneously because the level to which a split propagates is unknown before
traversal.

The \olcbskip is slightly slower than the \olcbtree\xspace on read-heavy
workloads because the \bskiplist may traverse horizontal chains at each level
before descending, visiting more nodes than a \btree's vertical descent. Under
OLC, every accessed node 
requires its own
version read and validation, incurring more atomic operations than in the \btree.

\begin{table}[t]
  \centering \resizebox{\columnwidth}{!}{
\begin{tabular}{@{}lcc@{}}
 \hline
\textit{System} & \textit{Throughput (txns/s)} & \makecell{\textit{Throughput relative}\\ \textit{to \obskipplot}} \\
 \hline
\obskipplot &5.77E4&1.00\\
\rwbskipplot&3.05E4&0.53\\
\follyplot&1.06E4&0.18\\
\rwbtreeplot&5.03E4&0.87\\
\olcbtreeplot&6.28E4&1.09\\
 \hline
\end{tabular}
}
\caption{Throughput (transactions/s) of the tested systems on the TPC-C
  benchmark (about 90\% reads).}
\label{tab:tpcc-throughput}
\end{table}

\para{TPC-C benchmark} \tabref{tpcc-throughput} shows that the \olcbskip
achieves $1.9\times$ and $5.4\times$ speedup over the \rwbskip and \folly, respectively, on the TPC-C benchmark. Furthermore, it achieves about
$1.1\times$ speedup over the \rwbtree\xspace and $0.9\times$ the throughput of the
\olcbtree. We found that the TPC-C benchmark contains about 90\% reads, which favors the \bplustree-based indices, as reflected in the results of the read-heavy YCSB workloads B and C.

\para{Deletes}
Overall, we find the \olcbskip achieves significantly higher delete throughput than the baselines on the delete-heavy workload. The \olcbtree\xspace does
not support deletion, so we omit it from this discussion.
For 50\%/50\% deletes/finds, the \olcbskip achieves $1.6\times$, $35\times$, and $116\times$ higher throughput than the \rwbskip, \folly, and \rwbtree, respectively. The \rwbtree suffers most under deletes because its OCC protocol relinquishes all locks and retries under write locks on each conflicting operation. For 5\%/95\% deletes/finds, the margins narrow to $0.9\times$, $4\times$, and $5\times$ the throughput of the \rwbskip, \folly, and \rwbtree, respectively. Just as in the YCSB workloads, the \olcbskip is well-suited to write-heavy workloads because the hybrid scheme only requires write locks starting from the promotion level.

\subsection{Latency.}

We report latency at 128 threads across four percentiles (50th, 90th, 99th, and
99.9th). \figref{combined_latencies} shows workloads A and C; we summarize the
remaining workloads in the text. Latency complements throughput
by revealing tail behavior: two systems with equal throughput may differ
substantially in latency variance.

Across all four percentiles, for the load phase, the \olcbskip achieves around
\latBskipL lower latency than the \rwbskip, \latFollyL lower than \folly, and \latRWBTreeL lower than the \rwbtree, but \latHigherOLCBTreeL higher latency than the \olcbtree. On point-operation
workloads (A, B, and C), it achieves around \latBskipABC lower latency than the \rwbskip, \latFollyABC lower than \folly, \latRWBTreeABC lower than the \rwbtree, and up to \latHigher higher latency than the
\olcbtree. On range-query workload E, it achieves around
\latBskipE lower latency than the \rwbskip, \latFollyE lower than \folly, and \latRWBTreeE lower than the \rwbtree.

Between workloads A and C, relative performance is similar except for the
\rwbtree. Its latency increases significantly above the 90th percentile on
workload A but remains competitive on workload C. The \rwbtree's OCC protocol
retries with write locks; as the insert fraction grows, more retries inflate
tail latency. The \bskiplists avoid this because the promotion level is known before
traversal, so retries require no harsher locking.

On zipfian workloads, we found that all systems have lower latencies than on
uniform workloads, and the relative performance between systems stays the same.

\subsection{Strong scaling.}

\begin{figure}[t]
  \centering
  \begin{tikzpicture}
    \begin{axis}[
        width=8cm, height=5cm,
        axis lines = left,
        xlabel = Num. Threads,
        ylabel style={align=center},
        ylabel = {Speedup},
        cycle list name=exotic,
        legend pos=north west,
        xmode=log,
        log basis x={2},
        legend columns=2,
        legend style={font=\small},
        ymax=100,
        ytick={20,40,60,80,100},
        log ticks with fixed point,
        xlabel shift={-2pt},
        xtick={1,2,4,8,16,32,64,128},
        xticklabels = {1,2,4,8,16,32,64,128},
      ]
      \addplot[mark=square*, color=o-bskip-dark, thick]
      coordinates{
        (1,   1)
        (2,   1.80)
        (4,   3.47)
        (8,   6.95)
        (16,  12.90)
        (32,  26.76)
        (64,  50.62)
        (128, 81.55)
      };
      \addlegendentry{\obskipplot}

    \addplot[mark=triangle*, color=bskip-dark, thick]
      coordinates{
        (1,   1)
        (2,   1.93)
        (4,   3.39)
        (8,   6.78)
        (16,  12.43)
        (32,  23.88)
        (64,  44.81)
        (128, 69.99)
      };
      \addlegendentry{\rwbskipplot}
      \addplot[mark=*, color=folly-dark, thick]
      coordinates{
        (1,   1)
        (2,   1.84)
        (4,   3.66)
        (8,   7.13)
        (16,  14.33)
        (32,  27.51)
        (64,  44.57)
        (128, 88.00)
      };
      \addlegendentry{\follyplot}

      \addplot[mark=diamond*, color=btree-dark, thick]
      coordinates{
        (1,   1)
        (2,   1.89)
        (4,   3.43)
        (8,   6.57)
        (16,  13.26)
        (32,  25.06)
        (64,  47.17)
        (128, 74.14)
      };
      \addlegendentry{\rwbtreeplot}

      \addplot[mark=pentagon*, color=olc-btree-dark]
      coordinates{
        (1,   1)
        (2,   1.96)
        (4,   3.20)
        (8,   6.04)
        (16,  12.30)
        (32,  23.46)
        (64,  43.60)
        (128, 64.01)
      };
      \addlegendentry{\olcbtreeplot}

    \end{axis}
  \end{tikzpicture}
  \caption{Strong scaling on uniform YCSB workload C (100\% finds).}
  \label{fig:scaling_c}
\end{figure}
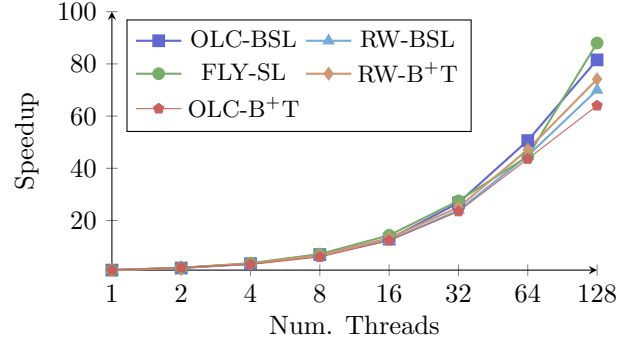

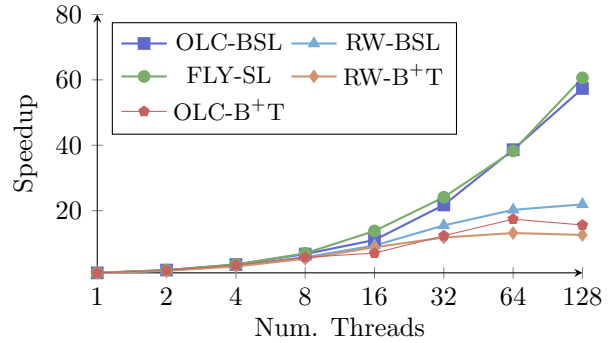
\begin{figure}[t]
  \centering
  \begin{tikzpicture}
    \begin{axis}[
        width=8cm, height=5cm,
        axis lines = left,
        xlabel = Num. Threads,
        ylabel style={align=center},
        ylabel = {Speedup},
        cycle list name=exotic,
        legend pos=north west,
        xmode=log,
        log basis x={2},
        legend columns=2,
        legend style={font=\small},
        ymax=80,
        ytick={20,40,60,80},
        log ticks with fixed point,
        xlabel shift={-2pt},
        xtick={1,2,4,8,16,32,64,128},
        xticklabels = {1,2,4,8,16,32,64,128},
      ]
      \addplot[mark=square*, color=o-bskip-dark, thick]
      coordinates{
        (1,   1)
        (2,   1.85)
        (4,   3.53)
        (8,   6.74)
        (16,  11.04)
        (32,  21.75)
        (64,  38.64)
        (128, 57.37)
      };
      \addlegendentry{\obskipplot}

    \addplot[mark=triangle*, color=bskip-dark, thick]
      coordinates{
        (1,   1)
        (2,   1.71)
        (4,   3.19)
        (8,   5.76)
        (16,  9.36)
        (32,  15.50)
        (64,  20.25)
        (128, 21.94)
      };
      \addlegendentry{\rwbskipplot}
      \addplot[mark=*, color=folly-dark, thick]
      coordinates{
        (1,   1)
        (2,   1.78)
        (4,   3.57)
        (8,   6.96)
        (16,  13.80)
        (32,  24.11)
        (64,  38.27)
        (128, 60.60)
      };
      \addlegendentry{\follyplot}

      \addplot[mark=diamond*, color=btree-dark, thick]
      coordinates{
        (1,   1)
        (2,   1.64)
        (4,   2.95)
        (8,   5.31)
        (16,  8.82)
        (32,  11.80)
        (64,  13.17)
        (128, 12.58)
      };
      \addlegendentry{\rwbtreeplot}

      \addplot[mark=pentagon*, color=olc-btree-dark]
      coordinates{
        (1,   1)
        (2,   1.76)
        (4,   3.40)
        (8,   5.74)
        (16,  7.03)
        (32,  12.34)
        (64,  17.41)
        (128, 15.62)
      };
      \addlegendentry{\olcbtreeplot}

    \end{axis}
  \end{tikzpicture}
  \caption{Strong scaling on the uniform YCSB load phase (100\% inserts).}
  \label{fig:scaling_load}
\end{figure}

Figures~\ref{fig:scaling_a}, \ref{fig:scaling_c}, and~\ref{fig:scaling_load} report scaling on uniform
workloads A and C, and the load phase, respectively, from 1 to 128 threads. All systems achieve speedup with
increasing thread count.

On workload A, the \olcbskip and \folly achieve around \olcbskipscalingA and
\follyscalingA speedups, respectively, while the \rwbskip, \rwbtree, and
\olcbtree\xspace lag behind at \rwbskipscalingA, \rwbtreescalingA, and
\olcbtreescalingA. The \olcbskip scales better because it eliminates the
memory-access bottlenecks of the \rwbskip and \rwbtree. On workload C, without insertions,
the gap closes and all systems achieve greater speedups: \olcbskipscalingC,
\rwbskipscalingC, \follyscalingC, \rwbtreescalingC, and \olcbtreescalingC for
the \olcbskip, \rwbskip, \folly, \rwbtree, and \olcbtree, respectively.

On the load phase, we find that the \olcbskip and \folly achieve
\olcbskipscalingL and \follyscalingL speedup, respectively, far ahead of the
\rwbskip, \rwbtree, and \olcbtree, which achieve only \rwbskipscalingL,
\rwbtreescalingL, and \olcbtreescalingL speedup, respectively. This scaling gap
comes from the cost of writing to shared node state in the different structures. The RW-lock-based structures write to the reader count in shared mode, invalidating the line in other cores' caches. The \olcbtree\xspace mitigates the issue by avoiding writes in shared mode, but splits eagerly and top-down: at every full node along an insert path, it write-locks that node and its parent, splits, and restarts the operation from the root. In contrast, the \olcbskip fixes its promotion height by an independent coin flip before the descent begins, so it write-locks only from the promotion level down and never restarts to split.




\section{Related work.}

We survey related work on optimistic concurrent \skiplists and blocked
\skiplists. \secref{intro} discusses existing OLC-based structures. 

\para{Optimistic concurrent \skiplists} The \skiplist's randomized structure
admits elegant lock-based and lock-free CC
protocols~\cite{herlihy2006provably, pugh1990concurrent,
  herlihy2007simple, fomitchev2004lock, fraser2004practical}.

The closest method is the optimistic \skiplist of Herlihy \emph{et
al.}~\cite{herlihy2006provably, herlihy2007simple}, which underlies
\folly~\cite{folly}. It traverses without acquiring locks and validates at the
target node. The \olcbskip inherits this simplicity while improving locality and
performance.

\para{Concurrent blocked \skiplists} Prior work improves the cache-friendliness
of concurrent \skiplists~\cite{xie2017parallelizing, na2023esl} but often
requires periodic restructuring of the upper levels, increasing tail latency.
ESL~\cite{na2023esl} inserts elements only into the bottom level and updates
upper index levels asynchronously. CSSL~\cite{SpreZeLe16} and
PI~\cite{xie2017parallelizing} periodically rebuild the upper levels, blocking
reads and writes during restructuring. S3~\cite{ZhangWuTa19} adaptively chooses
guard entries with a neural model, abandoning the \skiplist's probabilistic
guarantees.

We instead preserve the original \bskiplist structure to retain its simplicity
and probabilistic guarantees.

\section{Conclusion.}
\balance

This paper presents the \obskipnospace, a scalable concurrent in-memory index that
combines the cache-friendly node layout of the \bskiplist with OLC. The \obskip
preserves the \skiplist's simple structure while eliminating the space and
cache-invalidation overheads of the partitioned RW locks in the \rwbased
\bskiplist.

Our evaluation shows that on read-heavy workloads, the \olcbskip achieves
higher throughput than the tested \skiplists and competitive throughput with the
tested \btrees. On write-heavy workloads, the \olcbskip achieves
\tpWrite higher throughput over all tested indices, making it
well-suited for write-heavy applications such as databases and key-value stores.

We plan to integrate the \olcbskip into key-value stores such as
RocksDB~\cite{rocksdb} and LevelDB~\cite{LevelDB} to measure end-to-end
application impact. The \bskiplist's simple design, strong theoretical
guarantees, and OLC-based CC make it an ideal replacement for traditional
\skiplists in modern key-value stores.

\clearpage
\balance
\bibliographystyle{siamplain}
\bibliography{sample}
\end{document}